\documentclass[aps,prl,preprint,superscriptaddress]{revtex4-1} %preprint twocolumn
\usepackage{textcomp} %by me
\usepackage{graphicx} %by me
\usepackage{epstopdf}
\usepackage{setspace}
\usepackage{multirow} %by me
\begin{document}

% Use the \preprint command to place your local institutional report
% number in the upper righthand corner of the title page in preprint mode.
% Multiple \preprint commands are allowed.
% Use the 'preprintnumbers' class option to override journal defaults
% to display numbers if necessary
%\preprint{}

%Title of paper
\title{Observation of strongly entangled photon pairs \\from a nanowire quantum dot}

%\doublespacing
%Collaboration name if desired (requires use of superscriptaddress
%option in \documentclass). \noaffiliation is required (may also be
%used with the \author command).
%\collaboration can be followed by \email, \homepage, \thanks as well.
%\collaboration{}
%\noaffiliation

\date{\today}

\author{Marijn A. M. Versteegh}
\altaffiliation{These authors contributed equally to this work.\\ Corresponding author's email address: marijn.versteegh@univie.ac.at\\}
%\email{marijn.versteegh@univie.ac.at}
\author{Michael E. Reimer}
\altaffiliation{These authors contributed equally to this work.\\ Corresponding author's email address: marijn.versteegh@univie.ac.at\\}
\author{Klaus~D.~J\"ons}
\altaffiliation{These authors contributed equally to this work.\\ Corresponding author's email address: marijn.versteegh@univie.ac.at\\}
%\altaffiliation{These authors contributed equally to this work}
\affiliation
{Kavli Institute of Nanoscience, Delft University of Technology, Lorentzweg 1, 2628CJ Delft, The Netherlands}
%\affiliation
%{These authors contributed equally to this work}
\author{Dan Dalacu}
\author{Philip J. Poole}
\affiliation
{National Research Council of Canada, Ottawa, Canada, K1A 0R6}
\author{Angelo Gulinatti}
\affiliation
{Politecnico di Milano, Dipartimento di Elettronica Informazione e Bioingegneria, piazza Leonardo da Vinci 32 - 20133 Milano, Italy}
\author{Andrea Giudice}
\affiliation
{Micro Photon Devices, Via Stradivari 4 - 39100 Bolzano, Italy}
\author{Val Zwiller}
\affiliation
{Kavli Institute of Nanoscience, Delft University of Technology, Lorentzweg 1, 2628CJ Delft, The Netherlands}
\begin{abstract}
A bright photon source that combines high-fidelity entanglement, on-demand generation, high extraction efficiency, directional and coherent emission, as well as position control at the nanoscale is required for implementing ambitious schemes in quantum information processing, such as that of a quantum repeater. Still, all of these properties have not yet been achieved in a single device. Semiconductor quantum dots embedded in nanowire waveguides potentially satisfy all of these requirements; however, although theoretically predicted, entanglement has not yet been demonstrated for a nanowire quantum dot. Here, we demonstrate a bright and coherent source of strongly entangled photon pairs from a position controlled nanowire quantum dot with a fidelity as high as $0.859\pm0.006$ and concurrence of $0.80\pm0.02$. The two-photon quantum state is modified via the nanowire shape. Our new nanoscale entangled photon source can be integrated at desired positions in a quantum photonic circuit, single electron devices and light emitting diodes.
\end{abstract}
%%%%%%%%%%%%%%%%%%%%%%%%%%%%%%%%%%%%%%%%%%%%%%%%%%%%%%%%%%%%%%%%%%%%%
%% Start the main part of the manuscript here.
%%%%%%%%%%%%%%%%%%%%%%%%%%%%%%%%%%%%%%%%%%%%%%%%%%%%%%%%%%%%%%%%%%%%%
%%%%%%%%%%%%%%%%%%%%%%%%%%%%%%%%%%%%%%%%%%%%%%%%%%%%%%%%%%%%%%%%%%%%%
%% \section{Introduction}
%%%%%%%%%%%%%%%%%%%%%%%%%%%%%%%%%%%%%%%%%%%%%%%%%%%%%%%%%%%%%%%%%%%%%

\maketitle

There are demanding requirements for an `ideal' entangled-photon source for implementing ambitious schemes in quantum information processing, such as that of a quantum repeater~\cite{Briegel.Dur.ea:1998}. The source should meet the following criteria: high brightness combined with high-fidelity entanglement~\cite{Kwiat.Waks.ea:1999}, on-demand generation~\cite{Muller.Bounouar.ea:2014}, high extraction efficiency~\cite{Claudon.Bleuse.ea:2010}, directional~\cite{Reimer.Bulgarini.ea:2012} and coherent emission~\cite{Ates.Ulrich.ea:2009a}, as well as position control at the nanoscale~\cite{Juska.Dimastrodonato.ea:2013}. It is extremely difficult to meet all of these requirements in a single device. Good candidates are semiconductor quantum dots embedded in nanowires.

The high refractive index of a nanowire waveguide around a quantum dot ensures that the emitted light is guided in the desired direction and a tapered end makes the light extraction very efficient~\cite{Claudon.Bleuse.ea:2010}. With such a design, efficient single-photon generation has been demonstrated from a single nanowire quantum dot~\cite{Reimer.Bulgarini.ea:2012}. In addition, the emission mode-profile was shown to be directional and Gaussian~\cite{Munsch.ea:2014,Bulgarini.Reimer.ea:2014}, a key requirement for efficient long distance quantum communication in well-established telecommunication technology. Nanowires can be controllably positioned in uniform arrays~\cite{Borgstrom.Immink.ea:2007,Dorenbos.Sasakura.ea:2010}, with the ability to independently control the dot size and waveguide shell around it~\cite{Dalacu.Mnaymneh.ea:2012}. Silicon segments and substrates can be included in the design~\cite{Hertenberger.Rudolph.ea:2010,Kang.Gao.ea:2011,Hocevar.Immink.ea:2012,Munshi.Dheeraj.ea:2014} and electrical contacts have been demonstrated on single nanowires for single electron devices \cite{Kouwen.Reimer.ea:2010}, light emitting diodes \cite{Minot.Kelkensberg.ea:2007}, as well as single-photon avalanche photodiodes~\cite{Bulgarini.Reimer.ea:2012}. A significant advantage of using nanowire waveguides for efficient light extraction over other existing approaches, such as optical microcavities~\cite{Dousse.Suffczynski.ea:2010}, is the broad frequency bandwidth of operation~\cite{Bleuse.Claudon.ea:2011}, which is needed for achieving bright entangled photon pair generation via the biexciton-exciton radiative cascade. This approach is especially advantageous for quantum dots emitting over a large spectral range and may also be implemented with II-VI quantum dots where the biexciton binding energy is very large ($>20$\,meV)~\cite{Akimov.Andrews.ea:2006}.

A key feature of nanowires with embedded quantum dots grown in the [111]-direction is that the fine-structure splitting is expected to vanish~\cite{Singh.Bester:2009}, which should result in excellent entangled photon emission via the biexciton-exciton radiative cascade~\cite{Benson.Santori.ea:2000}. Our measurements realize this prediction and demonstrate the generation of strongly entangled photon pairs for the first time from a nanowire quantum dot. Our sources are ready to implement in advanced quantum information processing schemes without the need for any post-growth manipulation~\cite{Trotta.ea:2014} or temporal post-selection~\cite{Huber.Weihs.ea:2014}. Temporal post-selection can be a major source of photon losses and puts additional requirements on the measurement, thus limiting the scalability of quantum dot based entangled photon sources. For practical applications it is therefore very useful that we can avoid temporal post-selection. Finally, due to the efficient waveguiding and the tapered end, which we created during the bottom-up growth of the nanowire, we measure a light extraction efficiency of 18$\pm$3\,$\%$ for the source. Importantly, due to a recent breakthrough in the nanowire growth~\cite{Dalacu.Mnaymneh.ea:2012}, this high efficiency is obtained while potentially meeting all of the criteria of an ideal entangled photon source.

\section{Results}
\subsection{Site-controlled quantum dots in tapered nanowire waveguides.}
\begin{figure}[h]%
\includegraphics*[width=0.70\linewidth]{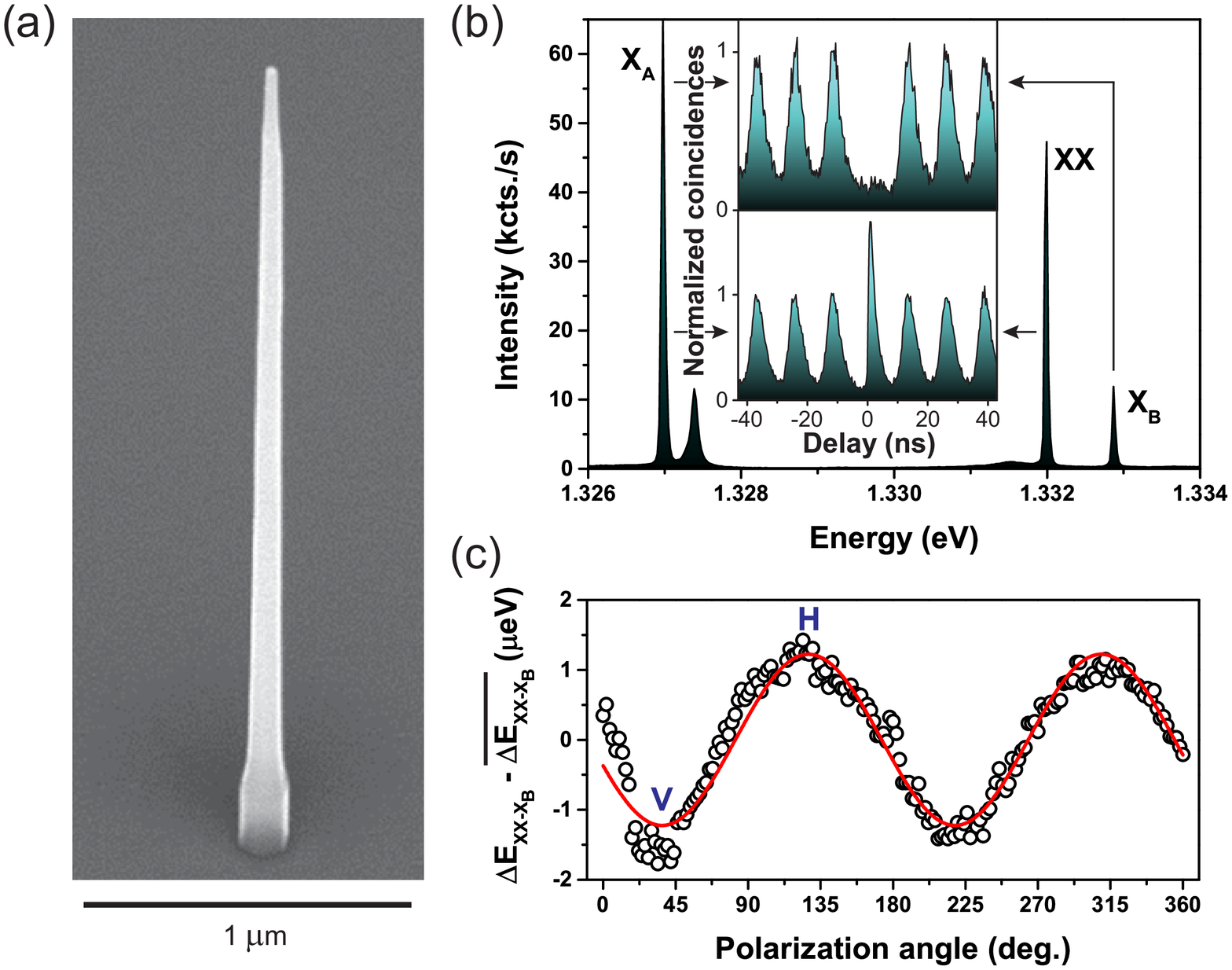}
\caption{\textbf{Nanowire quantum dot sample}. (a) Scanning electron microscopy image of a tapered nanowire waveguide with embedded quantum dot. (b) Photoluminescence spectrum of a single InAsP quantum dot embedded in an InP nanowire. The spectrum was taken at the excitation power used for the cross-correlation measurements needed to reconstruct the density matrix (100\,nW), which is close to saturation of both XX and X$_B$ transitions. Note that the excitonic transition X$_A$ saturates the CCD camera. (c) Polarization-dependent measurement to determine the excitonic fine-structure splitting. To increase the accuracy of the polarization measurement we plot the relative difference between biexciton XX and exciton X$_B$ emission energy. The amplitude of the sine-function fit indicates a fine-structure splitting of 1.2\,\textmu{}eV.
  }
\label{fig:fig1}
\end{figure}

The nanowires were grown by selective-area chemical beam epitaxy, which allows for control of the dot size and position, as well as enabling growth of the waveguide shell around the dot for efficient light extraction (see Methods). This technique has been demonstrated to yield defect-free, pure wurtzite nanowires, which is essential to obtain long single-photon coherence~\cite{Reimer14.ea:2014}. Fig.~\ref{fig:fig1}a shows a scanning electron microscopy image of a tapered InP nanowire waveguide containing an InAsP segment, 200~nm from the nanowire base, defining the optically active quantum dot that we study.

A spectrum taken under the excitation condition used for the quantum state tomography measurements is depicted in Fig.~\ref{fig:fig1}b. By performing cross-correlation measurements ~\cite{Baier.Malko.ea:2006}, shown in the inset, and power-dependent measurements (see Supplementary Fig.~1), we identified the biexciton (XX) and exciton (X$_A$ and X$_B$) transitions. The XX-X$_B$ cascade produces entangled photons. In contrast, a weak cascade is observed for XX-X$_A$ which does not show entanglement. From these observations, X$_A$ could be either a charged exciton or an exciton with a different hole state than X$_B$ as is permitted by the wurtzite crystal structure. The transitions XX and X$_B$ are resolution limited, single-photon interference measurements show excellent coherence of our entangled photons (see Supplementary Fig.~2). Autocorrelation measurements at saturation of the XX and X$_B$ transitions show strong antibunching, indicative of nearly perfect single photon pairs from the XX-X$_B$ cascade (see Supplementary Fig.~3).

\subsection{Light extraction efficiency.}
From the single-photon detector counts, 55 kilocounts per second for XX and 15 kilocounts per second for X$_B$, under pulsed excitation at 80\,MHz we calculate a collection of 7.9 million XX and 2.0 million X$_B$ photons per second into the first objective when taking our $0.7\pm0.1$\,\% setup efficiency at $\sim$900\,nm into account. The X$_A$ transition shows the highest intensity of 12.1 million photons per second and saturates our spectrometer's CCD camera with integration times as short as 1\,s under the excitation conditions used throughout our study (Fig~\ref{fig:fig1}b). This radiative recombination pathway from XX competes strongly with the X$_B$ emission, thus reducing the entangled photon pair generation efficiency. Taking into account this competing recombination pathway for XX, we calculate a light extraction efficiency of 18$\pm$3\,$\%$ for the source. We expect that adding a gold mirror with thin dielectric layer below the nanowire will boost the efficiency nearly two-fold~\cite{Bleuse.Claudon.ea:2011}. Combining this mirror with further engineering of the nanowire shape promises extraction efficiencies exceeding 90\,$\%$~\cite{Friedler09.ea:2009}.

\subsection{Low fine-structure splitting system.}
Using polarization-dependent measurements, presented in Fig.~\ref{fig:fig1}c, we obtain an estimation for the excitonic fine-structure splitting, $S$, by subtracting the XX transition from the X$_B$ transition energy~\cite{Young.Stevenson.ea:2009}. We obtain from the sine-function fit a fine-structure splitting of 1.2\,\textmu{}eV. In the case of nanowires, the small fine-structure splitting is a result of growth on a [111]-oriented substrate and the symmetric hexagonal cross-section of the nanowire core, defining the quantum dot. This small value that we measure for the fine-structure splitting is crucial for the entanglement observation between XX and X$_B$ photons without any temporal post-selection and is representative of the sample where, remarkably, a high percentage ($>50$\%) of the measured quantum dots show a fine-structure splitting below 2\,\textmu{}eV (Supplementary Fig.~4). The period of precession of the X$_B$ spin can be estimated as $h/S=3.5$\,ns~\cite{Stevenson.Hudson.ea:2008}, where $h$ is Planck's constant. This period of precession is much longer than the X$_B$ lifetime of $0.50 \pm 0.01$\,ns as extracted from the XX-X$_B$ cross-correlation measurements without polarization selection (Supplementary Fig.~5). Therefore, the X$_B$ spin precession has only little influence on the correlations in polarization between the two photons.

\subsection{Polarization-entangled photon pairs.}
\begin{figure}[h]%
\includegraphics*[width=0.70\linewidth]{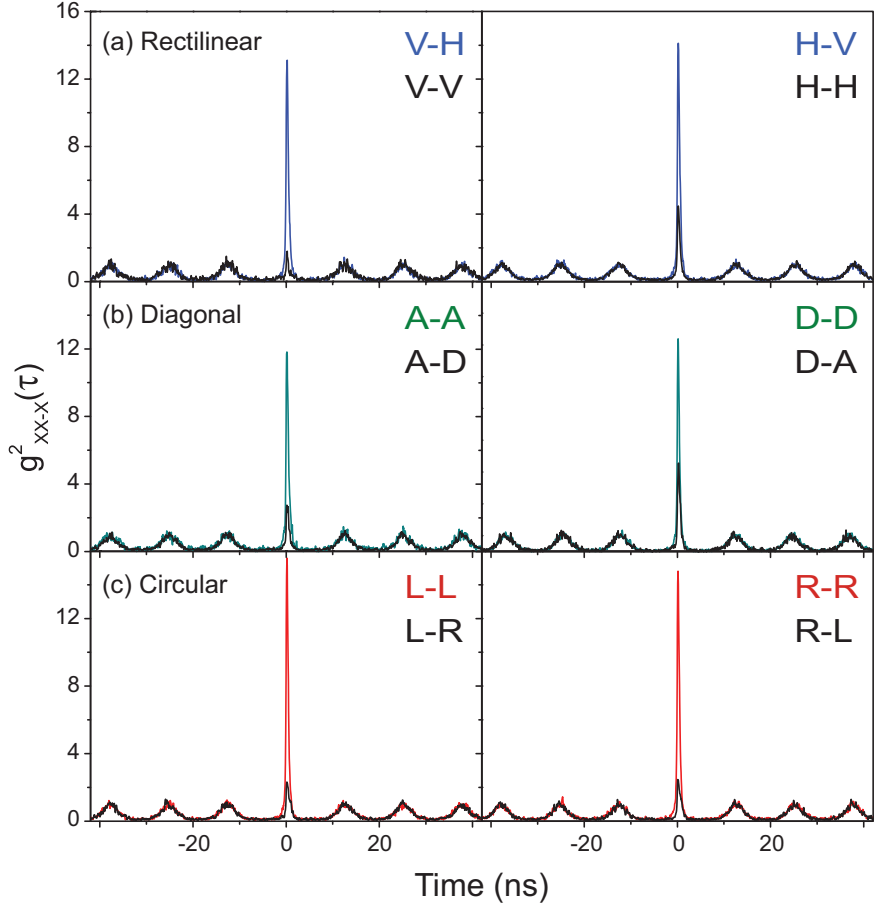}
\caption{\textbf{Cross-correlation measurements for the three different bases:} (a) rectilinear, (b) diagonal, and (c) circular. The plotted data is normalized to the Poisson level of the side peaks. Start: biexciton; stop: exciton X$_B$. The first letter stands for the measured polarization of the biexciton photon, whereas the second letter stands for the measured polarization of the exciton photon.
}
\label{fig:fig2}
\end{figure}

Twelve cross-correlation measurements in the rectilinear, diagonal, and circular polarization bases are shown in Fig.~\ref{fig:fig2}, where each histogram is composed of 64\,ps time bins. In the correlation measurements, the first letter stands for the measured polarization of the XX photon whereas the second letter stands for the X$_B$ photon. The strong correlations in \emph{HV, VH, DD, AA, RR}, and \emph{LL}, together with the weak correlations in \emph{HH, VV, DA, AD, RL}, and \emph{LR}, show that the two photons from the XX-X$_B$ cascade are entangled. Here, $H$ and $V$ are orthogonal linear polarizations (horizontal and vertical), $D=\left(H+V\right)/\sqrt{2}$ and $A=\left(H-V\right)/\sqrt{2}$ are diagonal and antidiagonal linear polarizations, whereas $R=\left(H+iV\right)/\sqrt2$ and $L=\left(H-iV\right)/\sqrt2$ are righthanded and lefthanded circular polarizations.

The quantum state we observe is different from the state that is measured for self-assembled quantum dots~\cite{Akopian.Lindner.ea:2006,Young.Stevenson.ea:2006,Hafenbrak.Ulrich.ea:2007,Juska.Dimastrodonato.ea:2013,Kuroda.Mano.ea:2013,Trotta.ea:2014}. Typically, one measures for the XX-X cascade bunching (positive correlations) in $HH$ and $RL$, and antibunching (negative correlations) in $HV$ and $RR$. However, we observe the opposite (Fig.~\ref{fig:fig2}a \& c). Only in the diagonal basis we see the usual correlations: bunching in $DD$ and antibunching in $DA$. These results show that the two-photon quantum state is closer to $(|HV\rangle+|VH\rangle)/\surd2$ than to the commonly measured state $(|HH\rangle+|VV\rangle)/\surd2$~\cite{Akopian.Lindner.ea:2006,Young.Stevenson.ea:2006,Hafenbrak.Ulrich.ea:2007,Juska.Dimastrodonato.ea:2013,Kuroda.Mano.ea:2013,Trotta.ea:2014}.

\subsection{Quantum state tomography.}
We performed quantum state tomography~\cite{James.Kwiat.ea:2001} to determine more precisely the quantum state of the photons and the degree of entanglement. The raw cross-correlation measurements needed to reconstruct the density matrix are shown in Supplementary Fig.~6. The resulting density matrix is given in Figs.~3a and 3b. The concurrence is $0.57\pm0.02$. A positive value for the concurrence means that the correlations cannot be explained classically and that the photons are quantum entangled. In this calculation all correlation counts in the full time window of 6.02\,ns are taken into account. The two-photon state has a fidelity of $0.762\pm0.002$ to the maximally entangled state $(|JJ\rangle+|WW\rangle)/\surd2$, where $J=H  e^{-i\beta}\cos\alpha+V e^{-i\beta}\sin\alpha$ and $W=-H  e^{i\beta}\sin\alpha+V e^{i\beta}\cos\alpha$ are two orthogonal elliptical polarizations. The angles $\alpha$
and $\beta$ are specified in Table 1. The classical limit is 0.5, so this result shows a strong degree of entanglement, even without temporal selection.

Selection of a narrower time window yields higher values for the concurrence and the fidelity (Table 1). For example, for a time window of 0.13 ns we calculate a concurrence of $0.80\pm0.02$ and a fidelity of $0.854\pm0.006$. The density matrix for this time window is presented in Supplementary Fig.~7. Temporal selection yields stronger entanglement, because within a narrow time window the effects of spin precession and dephasing processes are smaller~\cite{Stevenson.Hudson.ea:2008}. When we do not restrict our analysis to states of the form $(|JJ\rangle+|WW\rangle)/\surd2$, but instead calculate the fidelity to a general maximally entangled two-photon state, we find only slightly higher values (Table 1). The maximally entangled states to which the fidelity is maximal are very close to states of the form $(|JJ\rangle+|WW\rangle)/\surd2$.

\subsection{Two-photon quantum state modified by birefringence.}
Why do we measure $(|JJ\rangle+|WW\rangle)/\surd2$ and not the usual two-photon state that is measured for quantum dots, namely $(|HH\rangle+|VV\rangle)/\surd2$? The most probable reason is that the nanowire has a small anisotropy: it could have a slightly elongated cross section. An extreme case is shown in the SEM image of Fig.~4b. Such an anisotropy may be formed during the growth of the cladding around the core, and would then be unrelated to the shape of the quantum dot (for details of the growth, see nanowire elongation in the Methods section). As a comparison, we show a symmetric nanowire waveguide in the SEM image of Fig.~4a. In case of an elongated cross section the effective refractive indices are different for the polarizations along the short and the long axis of the nanowire. Here, we could imagine that the quantum dot emits photon pairs in the usual entangled quantum state $(|HH\rangle+|VV\rangle)/\surd2$. As the emitted photons are guided along the nanowire, the two-photon state is modified by birefringence into $(|JJ\rangle+|WW\rangle)/\surd2$, as is illustrated in Fig. 3c. Thus, $HH$ and $VV$ correlations rotate into predominantly $RR$ and $LL$ correlations, while $RL$ and $LR$ turn mostly into $HV$ and $VH$, which explains the observations of Fig.~\ref{fig:fig2}. For a nanowire waveguide of 6 $\mu$m length a difference of effective refractive index of order 0.1 would be enough to explain the magnitude of the observed rotation. Apart from birefringence in the waveguide, the polarization state of the emitted photons could also have been influenced by $\Gamma_7$ and $\Gamma_9$ hole mixing in the wurtzite quantum dot.

\section{Discussion}
In summary, we used a wurtzite nanowire quantum dot to generate single pairs of polarization-entangled photons with a fidelity as high as $0.859\pm0.006$ and a concurrence up to $0.80\pm0.02$. Furthermore, a high degree of entanglement is maintained (fidelity of $0.762\pm0.002$) without any temporal post-selection. This first observation of entangled photon pair generation from a nanowire quantum dot, which combines the desired properties of an ideal entangled photon source, opens new opportunities in quantum optics, integrated quantum photonic circuits~\cite{Politi.Cryan.ea:2008,Silverstone.Bonneau.ea:2014} and quantum information processing.

To realize an ideal entangled photon source in future work there are several properties of our source to consider. First, quantum dot entangled photon sources have not yet reached the fidelity or concurrence values of parametric down-conversion sources~\cite{Kwiat.Waks.ea:1999,Fedrizzi:2007}. However, with recently available post-growth tuning methods to bring the fine-structure splitting of almost any quantum dot near zero~\cite{Trotta.ea:2014} and two-photon resonant excitation~\cite{Muller.Bounouar.ea:2014}, the fidelity of these quantum dot sources are approaching that of parametric down-conversion sources. Second, the single-photon coherence of the emitted photon pairs is not yet Fourier-transform limited, which is needed for advanced quantum information processing schemes. Such Fourier-transform limited photons may be reached by combining two-photon resonant excitation techniques~\cite{Muller.Bounouar.ea:2014, Jayakumar:2013}, cooling of the quantum dot sample to 300\,mK~\cite{Reimer14.ea:2014}, and by accelerating the quantum dot emission via the Purcell effect~\cite{Dousse.Suffczynski.ea:2010}. Finally, the major advantage of tapered nanowire waveguides over other approaches is the light extraction efficiency, which promises entangled photon pair extraction efficiencies exceeding 90\,$\%$ due to the broadband frequency of operation~\cite{Friedler09.ea:2009}. Such efficiencies would surpass the state-of-the-art entangled photon pair efficiency of 12\,$\%$~\cite{Dousse.Suffczynski.ea:2010}, without the stringent requirements needed to engineer both the exciton and biexciton into resonance with a cavity mode by using post-growth manipulation of pre-selected quantum dots.

%%%%%%%%%%%%%%%%%%%%%%%%%%%%%%%%%%%%%%%%%%%%%%%%%%%%%%%%%%%%%%%%%%%%%
%% \section{Conclusion}
%%%%%%%%%%%%%%%%%%%%%%%%%%%%%%%%%%%%%%%%%%%%%%%%%%%%%%%%%%%%%%%%%%%%%

\section{methods}
\subsection{Nanowire quantum dot growth}
The InP nanowires containing single InAsP quantum dots were grown using chemical beam epitaxy (CBE) with trimethylindium (TMI) and pre-cracked PH$_3$ and AsH$_3$ sources. The nanowires were grown on a SiO$_2$-patterned (111)B InP substrate consisting of circular holes opened up in the oxide mask using electron-beam lithography and a hydrofluoric acid wet-etch. Au was deposited  in these holes using a self-aligned lift-off process, which allows the nanowires to be positioned at known locations on the substrate~\cite{Dalacu.Kam.ea:2009}. The thickness of the deposited gold is chosen to give 20\,nm diameter particles. The nanowires were grown at $420^\circ$C with a TMI flux equivalent to that used for a planar InP growth rate of $0.1\mu$m/hr on $(001)$ InP substrates at a temperature of $500^\circ$C. The growth is a two step process: ($i$) growth of a nanowire core containing the quantum dot, nominally 200\,nm from the nanowire base, and ($ii$) cladding of the core to realize nanowire diameters for efficient light extraction (around 200\,nm). The quantum dot diameters are determined by the size of the nanowire core. In this study, we investigated quantum dot diameters ranging from $\sim$25-30\,nm.

The nanowire core was grown for 26 minutes at a PH$_3$ flow of 3\,sccm. The dot was incorporated by switching from a PH$_3$ to an AsH$_3$ overpressure for 3 seconds after 15 minutes of growth. This growth time results in a quantum dot height of $\sim$6\,nm as determined in our previous studies~\cite{Dalacu.Mnaymneh.ea:2012}, using an energy-dispersive X-ray spectroscopy line scan along the nanowire, for a sample with nominally identical growth conditions. We note that our quantum dots are grown during 3\,s, resulting in taller quantum dots with longer emission wavelength, as compared to the work of Huber et al.~\cite{Huber.Weihs.ea:2014} who used a growth time of 2\,s. Our growth conditions result in very small fine-structure splittings as shown in Supplementary Fig.~4.

The nanowire cladding was grown by increasing the PH$_3$ flow to 9\,sccm. The total growth time was 120 minutes. To realize the smooth tapering towards the tip, the nanowire was made longer than the diffusion length of indium. Most nanowires, including the quantum dots, have a pure wurtzite crystal structure. The nanowire axis is the wurtzite $c$-axis.

\subsection{Nanowire waveguide elongation}

Axial growth is nominally constrained by the oxide opening and the nanowire cross-section has a hexagonal symmetry. Optimal coupling of the quantum dot emission to the waveguide mode requires diameters exceeding that of the oxide opening. This is achieved by increasing the axial growth time, which results in the nanowire overgrowing the oxide opening. No longer constrained by the opening, the hexagonal symmetry may be distorted (see Fig.~4). This asymmetry results in a geometric birefringence and concomitant rotation of the polarization state emitted by the quantum dot.
%(for details see Supplementary Note 1)

\subsection{Optical measurements}
The optical measurements were performed in a standard confocal microscopy setup where the quantum dot sample is cooled to a temperature of T\,=\,5\,K in a closed-cycle cryostat. The setup consists of two spectrometers both equipped with red-enhanced single photon avalanche diodes having 75\,ps time resolution, dark count rates as low as 80 counts per second and quantum efficiency of 11.5\,\% at 930\,nm~\cite{Gulinatti.Rech.ea:2012}.
A set of waveplates and polarizers placed in front of each spectrometer was used to perform polarization-dependent cross-correlation measurements. One spectrometer is set to the biexciton~(XX) transition and the other to the exciton~(X$_B$) transition. Each correlation measurement was done with 6000\,s of integration to reach over 1000 correlations in each side peak. For all photoluminescence and correlation measurements we use a Ti:Sapphire laser emitting 3\,ps long pulses at 750\,nm with a repetition rate of 80\,MHz to excite the quantum dot.

%%%%%%%%%%%%%%%%%%%%%%%%%%%%%%%%%%%%%%%%%%%%%%%%%%%%%%%%%%%%%%%%%%%%%
%% The "Acknowledgement" section can be given in all manuscript
%% classes.  This should be given within the "acknowledgement"
%% environment, which will make the correct section or running title.
%%%%%%%%%%%%%%%%%%%%%%%%%%%%%%%%%%%%%%%%%%%%%%%%%%%%%%%%%%%%%%%%%%%%%
\begin{acknowledgments}
\section{Acknowledgement}
We acknowledge S. H\"ofling, T. Braun, and R. Trotta for scientific discussions, L. Schweickert for the image illustrating the nanowire birefringence. This research was supported by the Dutch Foundation for Fundamental Research on Matter (FOM projectruimte 12PR2994), ERC, and the European Union Seventh Framework Programme 209 (FP7/2007-2013) under Grant Agreement No. 601126 210 (HANAS).

\end{acknowledgments}

\section{Author contributions}
M.E.R., M.A.M.V., and V.Z. conceived and designed the experiments.
K.D.J., M.E.R., and M.A.M.V. performed the experiments. D.D. and P.J.P. fabricated the sample. M.A.M.V., M.E.R., and K.D.J. analyzed the data. A.Gu. and A.Gi. developed the detectors. M.E.R., M.A.M.V., K.D.J., and V.Z. wrote the manuscript with input from the other authors.

\section{Competing financial interests}
The authors declare that they have no competing financial interests.

%%%%%%%%%%%%%%%%%%%%%%%%%%%%%%%%%%%%%%%%%%%%%%%%%%%%%%%%%%%%%%%%%%%%%
%% The same is true for Supporting Information, which should use the
%% suppinfo environment.
%%%%%%%%%%%%%%%%%%%%%%%%%%%%%%%%%%%%%%%%%%%%%%%%%%%%%%%%%%%%%%%%%%%%%
%\begin{suppinfo}
%
%This will usually read something like: ``Experimental procedures and
%characterization data for all new compounds. The class will
%automatically add a sentence pointing to the information on-line:
%
%\end{suppinfo}

%%%%%%%%%%%%%%%%%%%%%%%%%%%%%%%%%%%%%%%%%%%%%%%%%%%%%%%%%%%%%%%%%%%%%
%% The appropriate \bibliography command should be placed here.
%% Notice that the class file automatically sets \bibliographystyle
%% and also names the section correctly.
%%%%%%%%%%%%%%%%%%%%%%%%%%%%%%%%%%%%%%%%%%%%%%%%%%%%%%%%%%%%%%%%%%%%%
%\bibliography{All}

\newpage

\begin{figure}[h]%
\includegraphics*[width=0.70\linewidth]{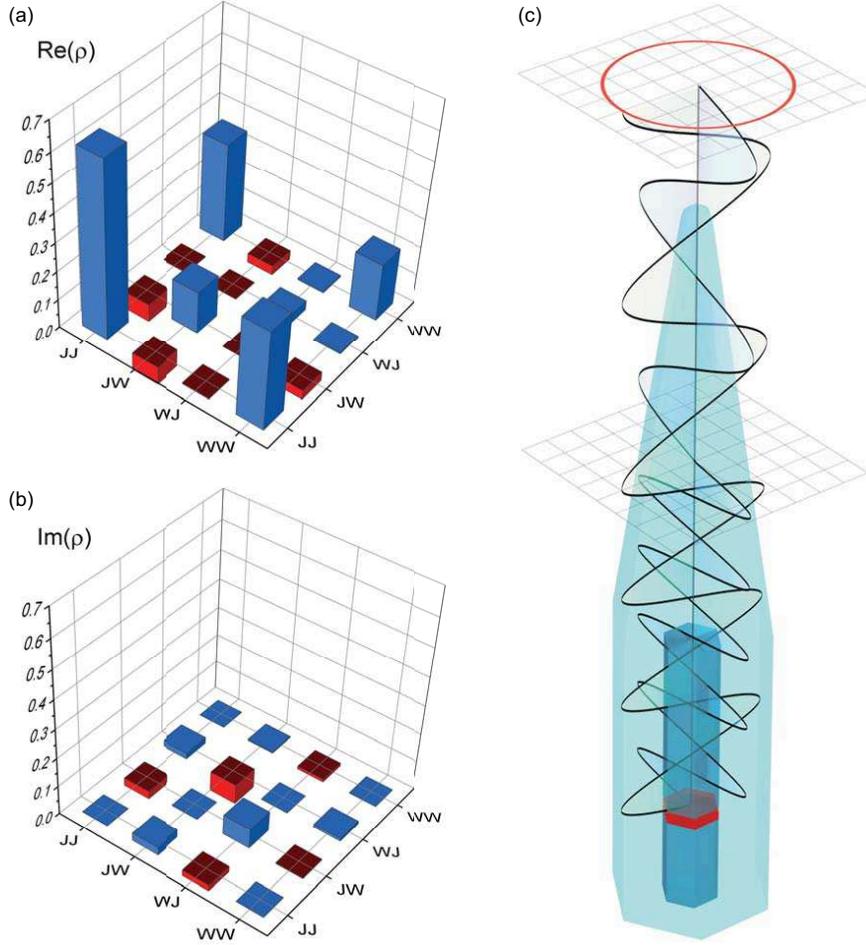}
\caption{\textbf{Quantum state tomography}. Real (a) and imaginary part (b) of the density matrix for the full time window of 6\,ns, in the rotated basis. The positive matrix elements are blue, and the negative matrix elements are red. (c) Illustration of the effect of birefringence in the nanowire. The orthogonal waves inside the nanowire experience different refractive indices, and therefore their wavelengths inside the waveguide are unequal. As a
result, the polarization of the light emission by the quantum dot (red) is modified leading to a different quantum state. The tapered section of the nanowire is more symmetric and is free of birefringence.
}
\label{fig:fig4}
\end{figure}

\begin{figure}[h]%
\includegraphics*[width=0.70\linewidth]{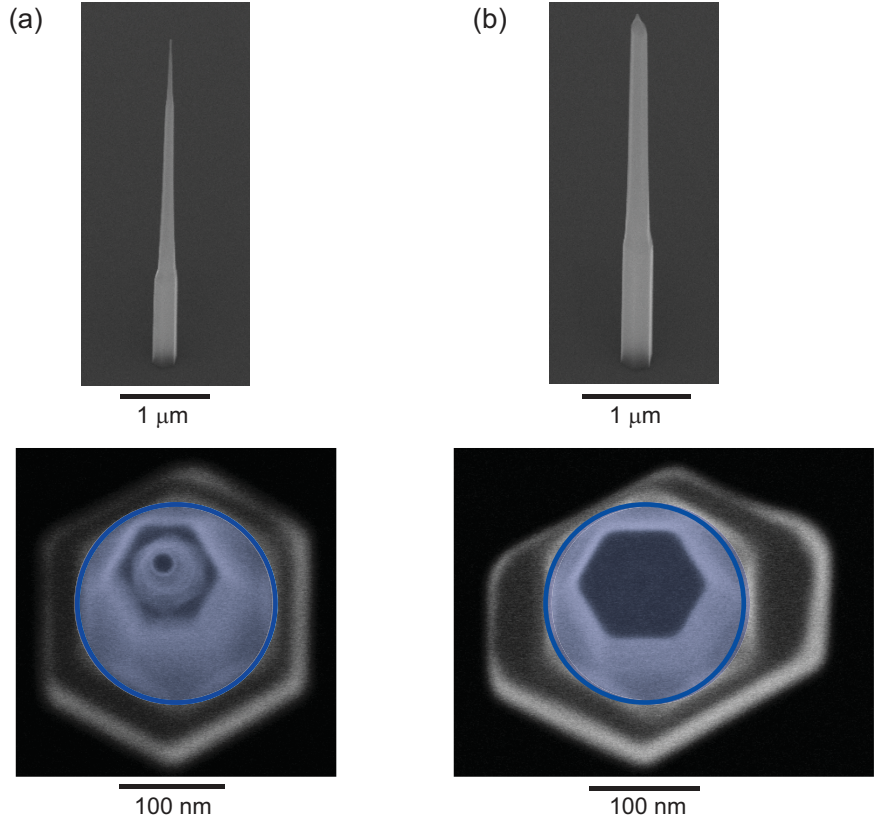}
\caption{\textbf{Nanowire birefringence}. SEM images of: (a) symmetric nanowire waveguide, and (b) asymmetric nanowire waveguide. Top panel: side-view SEM images of nanowires with tilt angle of 45 degrees. Bottom panel: SEM images of the nanowires viewed from the top at a small tilt angle. The blue shaded circle represents the opening in the SiO2 mask. The example of the nanowire elongation in (b) is an extreme example that leads to geometric birefringence and corresponding rotation of the quantum state.
}
\label{fig:fig5}
\end{figure}

\clearpage
\begin{table}[t]%
\includegraphics*[width=0.9\linewidth]{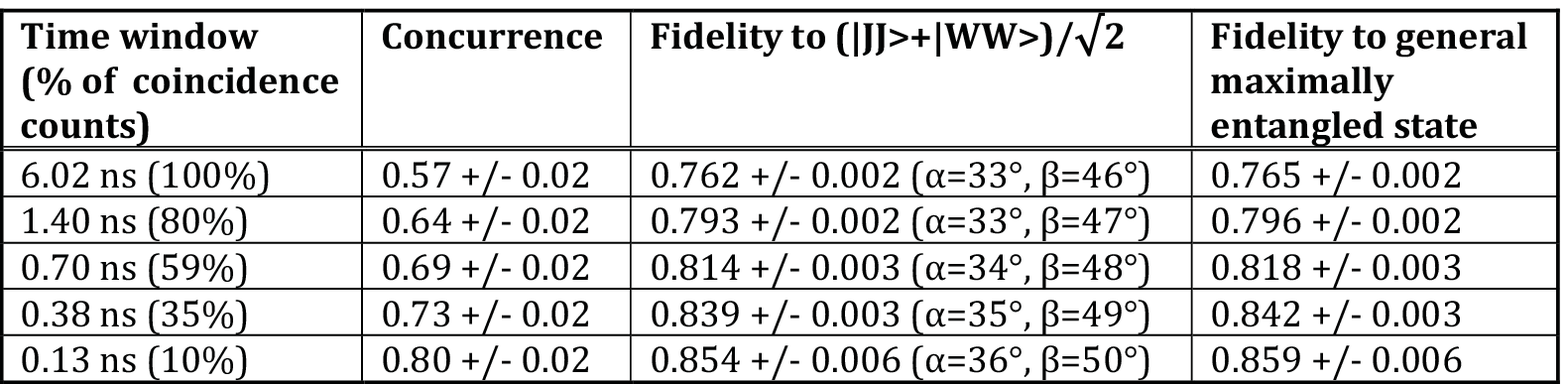}
\caption{Calculated concurrences and the fidelities for five different time windows. The percentage of the correlation events taken into account for a certain time window are given in brackets in the first column.}
\label{table1}
\end{table}
% The percentages in the first column stand for the amount of correlation events taken into account at this certain time window.

\end{document}